# Entanglement production and convergence properties of the variational quantum eigensolver

Andreas J. C. Woitzik[,1,*] Panagiotis Kl. Barkoutsos,[2] Filip Wudarski,[1,3,4,5] Andreas Buchleitner,[1,6] and Ivano Tavernelli[2]

[1]*Physikalisches Institut, Albert-Ludwigs-Universität Freiburg, Hermann-Herder-Straße 3, D-79104 Freiburg im Breisgau, Federal Republic of Germany*
[2]*IBM Research Europe GmbH, Zurich Research Laboratory, Säumerstrasse 4, 8803 Rüschlikon, Switzerland*
[3]*Institute of Physics, Faculty of Physics, Astronomy and Informatics, Nicolaus Copernicus University, Grudziądzka 5/7, 87-100 Toruń, Poland*
[4]*Quantum Artificial Intelligence Laboratory, Exploration Technology Directorate, NASA Ames Research Center, Moffett Field, California 94035, USA*
[5]*USRA Research Institute for Advanced Computer Science, Mountain View, California 94043, USA*
[6]*EUCOR Centre for Quantum Science and Quantum Computing, Albert-Ludwigs-Universität Freiburg, Hermann-Herder-Straße 3, D-79104 Freiburg im Breisgau, Federal Republic of Germany*

We perform a systematic investigation of variational forms (wave-function *Ansätze*), to determine the ground-state energies and properties of two-dimensional model fermionic systems on triangular lattices (with and without periodic boundary conditions), using the variational quantum eigensolver (VQE) algorithm. In particular, we focus on the nature of the entangler blocks which provide the most efficient convergence to the system ground state inasmuch as they use the minimal number of gate operations, which is key for the implementation of this algorithm in noisy intermediate-scale quantum computers. Using the concurrence measure, the amount of entanglement of the register qubits is monitored during the entire optimization process, illuminating its role in determining the efficiency of the convergence. Finally, we investigate the scaling of the VQE circuit depth as a function of the desired energy accuracy. We show that the number of gates required to reach a solution within an error $\varepsilon$ follows the Solovay-Kitaev scaling, $O[\log_{10}^c(1/\varepsilon)]$, with an exponent $c = 1.31 \pm 0.13$.

## I. INTRODUCTION

In the last decade we have experienced tremendous progress in quantum computing technologies [1–8]. A plethora of competing experimental realizations of quantum hardware has emerged and triggered the exploration of a new generation of quantum algorithms [9–14], in particular with the aim to gain novel insight into many-body physics or the electronic structure of atoms and molecules, and to enhance classical optimization strategies [15–26]. Despite all this progress current devices are still far from being fault tolerant, and exhibit limited connectivity, readout and gate errors, and short coherence times. Therefore, we are still confined to proof-of-principle studies using noisy intermediate-scale quantum (NISQ) [27] computers characterized by low depth circuits.

In this NISQ era of limited computational capabilities hybrid quantum-classical algorithms play a central role in the development of quantum computing applications. Some of the most promising approaches focus on the variational quantum algorithms (VQAs), which exploit the sampling from a parametrized quantum circuit of relatively low depth, i.e., the number of consecutive gates, and updating their parameters in an iterative process through a classical optimization scheme. The VQA aims at finding a near optimal solution of a given cost function, that can represent a physical Hamiltonian or combinatorial optimization problem. Two main algorithms have attracted considerable attention from the community— the quantum approximate optimization algorithm (and its extension the quantum alternate optimization ansatz) [28,29] and the variational quantum eigensolver (VQE) [30]. So far, research has focused on understanding and improving both the classical and quantum part of the algorithms as well as identifying suitable problems for their applications [31–43]. Worth mentioning are small molecular systems [17,19,20,44,45], while the treatment of larger problems is still hampered by NISQ imperfections (finite coherence times, insufficient gate fidelity, and readout errors). Therefore, an improved understanding of the operational properties of VQAs is still required in order to allow for near-to-optimal performance, i.e., for satisfactory convergence despite the device's restrictions.

A first step in this direction was taken by the proposal of a hardware-efficient VQE that exploits the available connections of a quantum device to parametrize the trial wave function for a molecular ground state, without significant increase of the overall circuit depth [20]. The main element of the quantum algorithm consists of a series of repeating blocks of single-qubit rotations and entangling gates, which need to sum up to less than a few hundred operations, in order to be executed within the limited coherence time of NISQ computers.

In our present paper, we address the versatility of the VQE paradigm on determining the ground state of simple,

---

*andreas.woitzik@physik.uni-freiburg.de

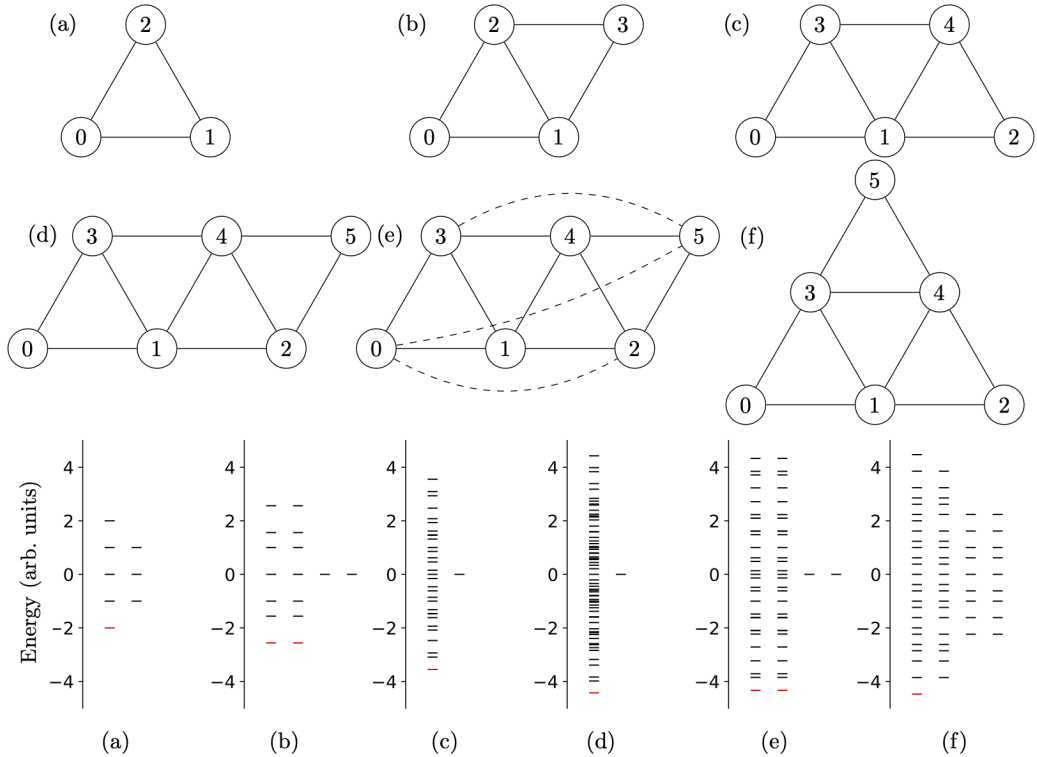

FIG. 1. Lattice models considered in this paper. Vertices represent sites that can be either occupied or unoccupied by spinless fermionic particles. Qubits are labeled by the sites they represent. Nearest-neighbor interactions occur along the edges. The dashed lines in (e) indicate interactions mediated by periodic boundary conditions. The spectra and state degeneracies of the corresponding Hamiltonians (a) $H_{\Delta_1}$, (b) $H_{\Delta_2}$, (c) $H_{\Delta_3}$, (d) $H_{\Delta_4^L}$, (e) $H_{\Delta_4^P}$, and (f) $H_{\Delta_4^S}$ are reported in the lower panel.

finite-size, noninteracting fermionic tight-binding models, as the elementary building block to assess ground-state properties of correlated electronic systems. These latter, highly complex, interacting many-particle systems remain a challenge for advanced computational solid state theory and therefore are an ideal target for potential improvements by quantum algorithmic elements. Specifically, we explore the optimal conditions for the application of the VQE to determine the ground state of the above simple lattice problems, focus on the design of scalable entangler blocks, and assess their efficiency by monitoring the convergence properties of the algorithm. We start by a comparison of the entanglement generated during the execution of the VQE on two distinct, isospectral three-qubit Hamiltonians exhibiting separable and entangled eigenstates, respectively. Subsequently we extend our analysis to larger two-dimensional (2D) tessellations with the above three-qubit plaquette as elementary unit. Finally, we investigate the scaling behavior of the VQE accuracy for variable numbers of optimization parameters. Our results are key to ponder whether VQE defines a viable strategy to deal with problems with a considerably larger number of degrees of freedom (e.g., with 50 to 100 qubits).

The paper is organized as follows: In Sec. II, we introduce the two elementary target-model Hamiltonians of interest, with their respective (non)separable ground states. Section III describes the architecture and gate structures of the VQE algorithm and elaborates on our numerical simulation procedures, as well as introducing the entanglement measures which we employ in the later analysis to monitor the VQE. Section IV investigates the entanglement generation upon execution and the speed of convergence, and assesses the scaling of the required computational resources during the optimization with the accuracy achieved upon convergence. In Sec. V we summarize the main results and give our outlook on the field.

## II. THE MODELS

For our analysis we elaborate on two different models which serve as target models for the VQE. We start from a simplified, noninteracting spinless Fermi-Hubbard model:

$$H = -t \sum_{\langle i,j \rangle} (c_j^\dagger c_i + c_i^\dagger c_j), \quad (1)$$

where nearest-neighbor sites $\langle i, j \rangle$ (on a 2D lattice to be specified) are coupled by a tunneling strength $t$, and $c_i^{(\dagger)}$ are the fermionic annihilation (creation) operators, respectively. In the following, all quantities will be measured in units of $t$, hence $t \equiv 1$.

### A. Fermionic Hamiltonians

First we restrict Eq. (1) to the case of three sites, which henceforth we call a *basic plaquette* [see Fig. 1(a)]. The direct mapping of this Hamiltonian to the qubit space is mediated by the Jordan-Wigner transformation [46], such that qubit states $|0\rangle$ and $|1\rangle$ are associated with unoccupied and occupied fermionic sites, respectively. Qubits are labeled by the lattice site they represent [47].

After applying the Jordan-Wigner transformation, Eq. (1) specialized to three sites turns into

$$H_{\Delta_1} = \tfrac{1}{2}(X_0 X_1 + Y_0 Y_1 + X_1 X_2 + Y_1 Y_2 + X_0 Z_1 X_2 + Y_0 Z_1 Y_2), \quad (2)$$

where $X_k$, $Y_k$, and $Z_k$ are Pauli matrices acting on the $k$th qubit ($k = 0, 1, 2$). We will refer to the Hamiltonian in Eq. (2), illustrated in Fig. 1(a), as $H_{\Delta_1}$, with ground-state energy $E_g = -2$ and associated eigenstate

$$\left|\Psi_g^{H_{\Delta_1}}\right\rangle = \frac{1}{\sqrt{3}}(|001\rangle + |010\rangle + |100\rangle). \quad (3)$$

The full spectrum of $H_{\Delta_1}$ reads

$$(-2, -1, -1, 0, 0, 1, 1, 2). \quad (4)$$

The ground state $|\Psi_g^{H_{\Delta_1}}\rangle$ is a particular example of an entangled state with nonzero concurrence but vanishing three-tangle (see Sec. III B). Starting from this basic plaquette unit, we extend our investigation to a series of larger lattices with four, five, and six sites [see Figs. 1(b)–1(f)]. In the following, we will refer to their corresponding Hamiltonians as $H_{\Delta_k}$, where $k$ indicates the number of basic plaquettes sharing one edge. For the case with six sites, we distinguish three nonequivalent arrangements: the linear $H_{\Delta_4^L}$ [Fig. 1(d)], the periodic (topologically equivalent to a ring) $H_{\Delta_4^P}$ [Fig. 1(e)], and the stacked $H_{\Delta_4^S}$ [Fig. 1(f)] forms. The corresponding energy spectra are depicted in the lower panel of Fig. 1.

### B. Separable Hamiltonian

In order to probe the role of entanglement generation on the efficiency of the computation, we also consider a second three-site Hamiltonian which shares the spectrum with $H_{\Delta_1}$ (4), while the corresponding eigenvectors are separable. Therefore, we call this Hamiltonian in the further analyses a *separable Hamiltonian*.

To derive the latter, we rotate the diagonal Hamiltonian $H_1 = \mathrm{diag}(-2, -1, -1, 0, 0, 1, 1, 2)$ in the computational basis

$$H_{\mathrm{sep}} = \left[Z \otimes \tfrac{1}{\sqrt{2}}(Z - X) \otimes X\right] H_1 \left[Z \otimes \tfrac{1}{\sqrt{2}}(Z - X) \otimes X\right]. \quad (5)$$

The ground state of this Hamiltonian is

$$\left|\Psi_g^{\mathrm{sep}}\right\rangle = \frac{1}{\sqrt{2}}(|001\rangle - |011\rangle) = |0\rangle \otimes |-\rangle \otimes |1\rangle, \quad (6)$$

where $|-\rangle = \frac{1}{\sqrt{2}}(|0\rangle - |1\rangle)$. This state is manifestly separable, and can be reached by applying local rotations to the qubits that encode each site, when starting in the separable initial state $|000\rangle$. The system described by (5) is therefore a good candidate for the discussion of the relevance of entanglement and its role for VQE-based optimization.

## III. METHODS

In this section we present different aspects of the hybrid quantum-classical VQE algorithm. First, we state the general formulation of the algorithm in Sec. III A. In Sec. III A 1 we discuss the parametrization of the wave function and properties related to the entangler blocks. Thereafter, we discuss the accuracy of solutions by the VQE in Sec. III A 2. In Sec. III B we present the entanglement measures we use. Finally, we give an overview of the parameters which are important for the analysis and how we tune them in Sec. III C.

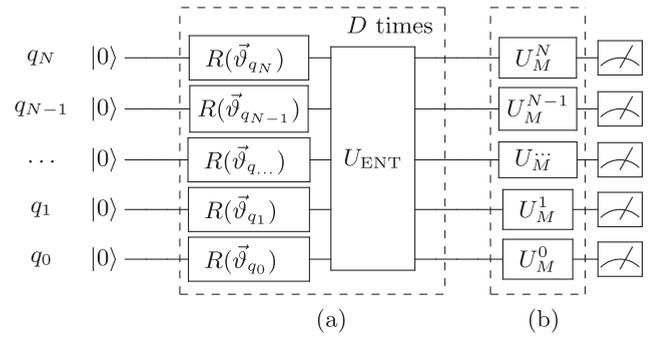

FIG. 2. Quantum circuit for the parametrization of the wave-function creation on an $N$-qubit system, where $\vec{\theta} = (\vec{\vartheta}_{q_0}, \ldots, \vec{\vartheta}_{q_{N-1}})$, and $\vec{\vartheta}_{q_i}$ describes the rotation angles for the single-qubit rotation of the $i$th qubit. The rotation angles are (in general) different for each of the $D$ blocks. (a) The repeating part of the circuit, consisting of single-qubit rotations $R(\vec{\vartheta})$ and entanglers $U_{\mathrm{ent}}$. The $D$ blocks define the full evolution operator $U(\vec{\theta})$ acting upon the wave function. (b) The premeasurement rotations generating the appropriate measurement basis.

### A. The VQE algorithm

The VQE is an algorithm that targets the minimum energy (ground-state energy) of a physical system represented by a Hamiltonian $H$. The operational basis for the VQE is the variational principle. Given a bounded Hamiltonian $H$, its expectation value with respect to a normalized wave function (vector) is always greater than or equal to the Hamiltonian's ground-state energy $E_g$ [48]:

$$\forall |\psi\rangle \in \mathcal{H}, \langle \psi | \psi \rangle = 1: \quad \langle \psi | H | \psi \rangle \geqslant E_g. \quad (7)$$

In the hybrid approach of the VQE, the variational optimization procedure is divided into two steps. The first one is performed by a quantum processing unit (QPU) and the second one is performed by a classical processing unit (CPU). The QPU is responsible for measuring the expectation values of Pauli operators with respect to the parametrized quantum state (the so-called trial wave function) that is constructed by the quantum circuit. Since real hardware often is restricted to a single measurement basis, the circuit contains premeasurement rotations ($U_M^i$ in Fig. 2) in order to allow the measurement in the Pauli basis. Later these Pauli expectation values help to infer the energy expectation value (see Sec. II A) and are passed to the CPU, where the new parameters are generated according to the classical optimization scheme. The new parameters are used to create an updated trial wave function, that is measured in the next iteration step. We repeat this process for a chosen number $\mathcal{I}$ of iterations.

*1. Quantum circuit structure and trial wave functions*

Now we scrutinize the quantum circuit of the VQE algorithm. Again, the goal of the algorithm is to create a quantum state which is close to the ground state, in order to measure an energy expectation value that is close to the ground-state energy. This is achieved by applying the quantum circuit to the initial state, which we set to $|0\rangle^{\otimes N}$. This state is evolved by the circuit $U(\vec{\theta})$ to the trial wave function $|\psi(\vec{\theta})\rangle$, which is then measured in some basis chosen by the premeasurement rotations $U_M$. The trial wave function is parametrized using a series of blocks built from single-qubit rotations $U_R(\vec{\theta}^k)$, followed by an entangler $U_{\text{ent}}$, that spans the required length of the qubit register. Since the single-qubit rotations are all local operations, $U_R(\vec{\theta}^k)$ can be written as a tensor product of the rotations of a single qubit:

$$U_R(\vec{\theta}^k) = \bigotimes_{i=0}^{N-1} R(\vec{\vartheta}_{q_i}^k), \tag{8}$$

where $R(\vec{\vartheta}_{q_i}^k)$ can be visualized as a rotation on the Bloch sphere of qubit $q_i$. We define

$$R(\vec{\vartheta}_{q_i}^k) = R_Z(\alpha_{q_i}^k) R_X(\beta_{q_i}^k) R_Z(\gamma_{q_i}^k). \tag{9}$$

This block sequence of single-qubit rotations and two-qubit entanglers is repeated for a variable number $D$ of times allowing more parameters for the optimization procedure. With this definition, the number of independent parameters is increasing as $3ND$ for an $N$-qubit system with $D$ blocks for the trial wave-function parametrization. The full unitary circuit operation is described by

$$U(\vec{\theta}) = \overbrace{U_{\text{ent}} U_R(\vec{\theta}^D) \ldots U_{\text{ent}} U_R(\vec{\theta}^1)}^{D \text{ times}}, \tag{10}$$

and the parametrized state is described by

$$|\psi(\vec{\theta})\rangle = U(\vec{\theta}) |0\rangle^{\otimes N}. \tag{11}$$

The quantum circuit corresponding to this unitary is depicted in Fig. 2. Note that the unitary $U(\vec{\theta})$ describes the full circuit, but not the premeasurement rotations. The nature of the entangler block can vary from case to case, and its purpose is to guarantee an efficient scan of the relevant part of the Hilbert space.

*2. Accuracy of the optimized solution*

One main issue in using the VQE algorithm to determine ground-state properties of quantum systems is related to the scaling of the error with the number of parameters included in the optimization process. In fact, a simple dimensional analysis shows that for an exhaustive sampling of the Hilbert space associated with a given quantum-mechanical problem one needs an exponentially large number of parameters. For example, for a system with $N$ qubits the dimensionality of the corresponding Hilbert space is $2^N$. However, the optimization of this exponentially large number of parameters using the hybrid VQE algorithm will frustrate the possibility to achieve any quantum advantage, since the optimization on the parameter space is still performed classically. Since we cannot sample the full Hilbert space exhaustively, it is crucial to choose a suitable subspace to sample from.

The Solovay-Kitaev (SK) theorem provides an upper bound for the number of gates (and therefore gate angles) required to achieve a desired accuracy for the energy. In short, the theorem states that for any target operation $U \in SU(2^N)$ there is a sequence $S = U_{s_1} U_{s_2} \ldots U_{s_D}$ of length $D = O[\log_{10}^c(1/\varepsilon)]$ in a dense subset of $SU(2^N)$ such that the error $d(U, S) < \varepsilon$, where $d(U, S) = \sup_{||\psi||=1} ||(U - S)\psi||$, and $U_{s_i}$ is the repeating unit in Eq. (10) (see also Fig. 2) with independent parameters $\vec{\theta}^{s_i}$. The theoretical worst-case upper bound of $c$ is 4 [49]. In our case, $U$ represents the $N$-qubit gate operation required to generate the exact ground-state wave function, while the set $S$ is represented by the parametrized sequence in Eq. (10). Although the subset of $SU(2^N)$ operations generated by the entangler blocks in Table I may not generate a dense subset of $SU(2^N)$ arbitrarily close to the exact unitary $U$ (the generator of the exact ground state), we analyze the convergence process numerically to find first indications of suitable entanglers for scaling (see Sec. IV C). We show a scaling relation between the VQE error and number $D$ of repeating blocks, for some models in Fig. 7.

The number $3ND$ of independent gate parameters (in an $N$-qubit system with $D$ blocks) is not the only variable playing a role in practical implementations. In fact, the precision with which we can set the gate angles in a quantum computing experiment is also limited by the available hardware and electronics. In this paper, we will investigate how these two factors, i.e., the number of degrees of freedom (independent gate parameters) and the decimal places (DP) of precision in setting the angles, affect the accuracy of the VQE energies. In particular, we will derive a scaling parameter $c$ for the case in which the distance $d$ above is replaced by $\varepsilon_e$, i.e., the error in the VQE ground-state energy.

**B. Entanglement measures**

Due to the rich structure of entanglement in multipartite systems, we limit our entanglement analysis to the basic plaquette, where one may distinguish quantum correlations within each possible pair of qubits $(i, j)$ (after tracing over the third qubit $k$) or within the entire system (tripartite entanglement). In order to quantify the amount of entanglement of the trial wave function along the optimization process, we define common entanglement measures. We use the general notion of concurrence as outlined in [50]. For a pure two-qubit state the concurrence is defined as [51]

$$C(|\Psi\rangle) = |\langle \Psi^* | \sigma_y \otimes \sigma_y | \Psi \rangle|, \tag{12}$$

where $\langle \Psi^* |$ is the transpose of $|\Psi\rangle$, in the standard basis $\{|00\rangle, |01\rangle, |10\rangle, |11\rangle\}$. To calculate the concurrence between two qubits of a higher-dimensional state, we need to calculate the concurrence for a mixed state $\rho_{i,j}$, which is derived after tracing over all qubits except $i$ and $j$ (in the case of the basic plaquette over qubit $k$). The concurrence is then given by the corresponding convex roof [52]:

$$C_{i,j} = C(\rho_{i,j}) := \inf_{p_n, |\Psi_n\rangle} \sum_n p_n C(|\Psi_n\rangle), \tag{13}$$

where $\rho_{i,j} = \sum_n p_n |\Psi_n\rangle \langle \Psi_n|, \quad \text{and} \quad p_n > 0. \tag{14}$

TABLE I. Comparison of 21 different entanglers built from gates typically implemented in present quantum machines [54]. The convergence speed is described by the fraction of altogether 1000 runs which lead to convergence with an error of not more than 2% of the ground-state energy. We use a three block circuit and the fermionic triangle Hamiltonian (2). We use a widely used gate notation, documented in Appendix A.

| ID | Circuit | Speed | ID | Circuit | Speed |
|---|---|---|---|---|---|
| 0 | $I$ | 0% | 1 | | 96.7% |
| 2 | | 98.8% | 3 | | 82.4% |
| 4 | | 84.1% | 5 | | 85.6% |
| 6 | | 0% | 7 | | 84.7% |
| 8 | | 62.3% | 9 | | 81.7% |
| 10 | | 95.9% | 11 | | 64.6% |
| 12 | | 64.2% | 13 | | 95.4% |
| 14 | | 98.8% | 15 | | 93.8% |
| 16 | | 70.8% | 17 | | 83.1% |
| 18 | $QFT$ | 34.9% | 19 | | 73.5% |
| 20 | | 74.2% | | | |

The concurrence can take values in the interval $C_{i,j} \in [0, 1]$, vanishes if and only if the state is separable, and equals 1 for maximally entangled states (e.g., Bell states). To quantify tripartite entanglement, we employ the measure of the three-tangle [53], which is defined as

$$\tau_3 \equiv \tau(i:j:k) := T_{j,k}^2 - \left(C_{i,j}^2 + C_{i,k}^2\right), \quad (15)$$

where $C_{i,j}$ is the concurrence between qubits $i$ and $j$, and

$$T_{j,k} := \sqrt{2 - 2\mathrm{Tr}(\rho_{j,k}^2)}. \quad (16)$$

The three-tangle measure is independent of the order of $i$, $j$, and $k$ and takes values in the interval [0,1]. In this paper, we propose to monitor the amount of entanglement in the VQE optimization process by integrating the concurrence and the three-tangle over the entire process. This amounts to summing up the entanglement levels (as measured by the concurrence and three-tangle) of the trial wave function at each iteration, according to

$$\bar{C}_{k,l} = \frac{1}{\mathcal{I}} \sum_{i=1}^{\mathcal{I}} C_{k,l}^{(i)}, \quad \bar{\tau}_3 = \frac{1}{\mathcal{I}} \sum_{i=1}^{\mathcal{I}} \tau_3^{(i)}, \quad (17)$$

where $\mathcal{I}$ is the number of iterations. For the given problem Hamiltonian $H_{\Delta_1}$, this measure indicates the amount of bipartite entanglement and three-tangle generated. We also investigate the amount of entanglement during the convergence process of the VQE for the separable Hamiltonian $H_{\mathrm{sep}}$. In this case, entanglement is not necessary for the convergence, but still affects the speed of the convergence.

### C. Simulations

All analyses presented in Sec. IV assume perfect conditions—no noise, no measurement errors, and high numerical precision, 15 orders of magnitude smaller than the minimal energy difference between two (nondegenerate) eigenstates of the Hamiltonians. Since we simulate the generation of the trial wave function, we can access all information on our system; in particular we can track how expectation values of energies or the amount of entanglement are changing throughout the convergence process.

In the simulations we use a stochastic direct search scheme—simultaneous perturbation stochastic approximation (SPSA)—that has proved to be suitable in hybrid scenarios [20], and set the SPSA parameters $\{\alpha, \gamma, c\} = \{0.602, 0.101, 0.01\}$ as reported in [20]. The SPSA algorithm needs a calibration process, which is performed before the actual VQE process. Because of the SPSA search scheme, the VQE—as we implement it—is a stochastic algorithm. Therefore, we need to evaluate the algorithm repeatedly in order to get proper statistics of the performance of the algorithm. We interchangeably call these repetitions of the algorithm *runs* or *repetitions*, which shall not be confused with the number $\mathcal{I}$ of iterations, which describes the number of trial wave functions generated in a single VQE optimization process. For different Hamiltonians we employ the algorithm with different numbers $\mathcal{I}$ of iterations (the number of trial wave-function measurements), calibration steps (the number of iterations to adjust SPSA parameters before running the VQE), repetitions (the number of random initializations of the full VQE cycle), and entangler blocks that are collected in Table II. Based on this setup, we quantify the fraction of instances which converge within a margin of 2% to the exact ground-state energy. We choose a 2% threshold for pragmatic reasons, as this threshold allows reasonable convergence rates within $\mathcal{I} = 1000$ iterations for the elementary, three-qubit plaquette.

TABLE II. Optimization parameters for different Hamiltonians. Tabulated parameters are as follows: $\mathcal{I}$, number of iterations of the VQE; NoC, number of calibrations to set the parameters for the SPSA optimization scheme; NoR, number of repetitions of the full algorithm; $D$, number of blocks (see Fig. 2).

| Hamiltonian | $\mathcal{I}$ | NoC | NoR | $D$ |
|---|---|---|---|---|
| $\Delta_1$ | 1000 | 100 | 1000 | 3 |
| $\Delta_2$ | 2000 | 200 | 500 | 5 |
| $\Delta_3$ | 4000 | 250 | 100 | 8 |
| $\Delta_4^S$ | 6000 | 300 | 100 | 12 |
| $\Delta_4^L$ | 6000 | 300 | 100 | 12 |
| $\Delta_4^P$ | 6000 | 300 | 100 | 12 |

## IV. RESULTS

### A. Efficiency of the entanglers: Speed of convergence

For the three-qubit Hamiltonian $H_{\Delta_1}$, Eq. (2), we investigate the speed of convergence for the 21 different entanglers listed in Table I. Based on our numerical experiments, we found that all of the entanglers acting upon the full qubit register allow convergence when the quantum circuit is composed of three or more blocks ($D \geqslant 3$). Furthermore, the speed of convergence depends on the number of blocks that compose the circuit. Figure 3 shows that an increased number $D$ of blocks leads to faster convergence of the algorithm in terms of the number $\mathcal{I}$ of iterations on the QPU.

An increase in the number of blocks leads to more single-qubit rotation angles to be optimized by the VQE algorithm. However, in a realistic scenario, hardware restrictions will limit the circuit depth due to rapidly growing errors (related to gate imperfections and limited coherence times). Another limiting factor for the increase in the number of blocks is the classical optimization. By increasing the number of parameters, the optimization loop executed by the CPU can hamper the overall performance. Therefore, one needs to find a compromise between the number of iterations required to converge, the number of parameters needed to be optimized, and the intrinsic hardware imperfections, which limit the amount of reliable quantum operations.

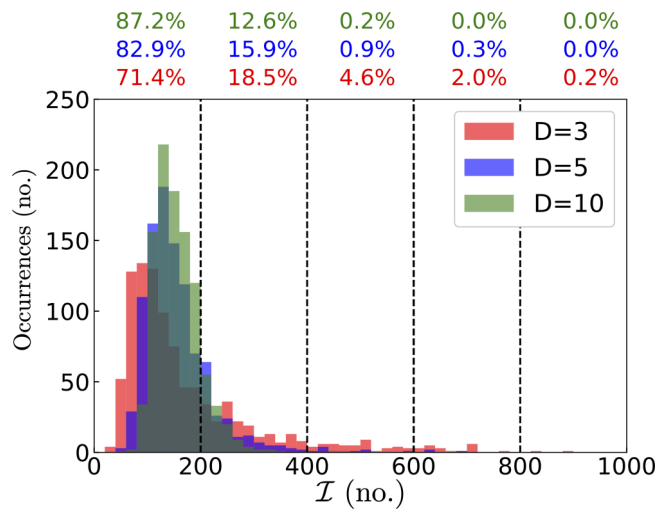

FIG. 3. Speed of convergence for different numbers $D$ of blocks (see legend) in the VQE quantum routine (see Fig. 2), when optimizing the ground-state energy of Hamiltonian $H_{\Delta_1}$ with the entangler 1 (see Table I). Convergence is here defined as the output energy of the routine matching the actual ground-state energy within an error margin of 2%, after not more than 1000 repetitions of the algorithm. Each simulation is done with 100 calibration steps and $\mathcal{I} = 1000$ iterations. The percentages on top of the plot (in the legend's color code) indicate the percentage of incidences of convergence within the first, second, etc., 200 iterations of, altogether, 1000 runs.

TABLE III. Percentage of runs (for number of runs see Table II) that lead to convergence with an error of not more than 2% of the ground-state energy. The table shows all types of entanglers (see Appendix B for their scaling structure) applied to the investigated Hamiltonians (see Table VII).

| Entangler | Percentage of convergence (%) | | | | |
|---|---|---|---|---|---|
| | $\Delta_2$ | $\Delta_3$ | $\Delta_4^L$ | $\Delta_4^P$ | $\Delta_4^S$ |
| Ent. 1 | 98.8 | 89 | 80 | 97 | 94 |
| Ent. 2 | 100 | 99 | 92 | 99 | 94 |
| Ent. 3 | 86.6 | 51 | 18 | 13 | 15 |
| Ent. 4 | 85.6 | 33 | 6 | 10 | 15 |
| Ent. 5 | 93.6 | 58 | 19 | 13 | 29 |

As three blocks proved sufficient for the convergence of all circuits with entanglers that span the full quantum register, we compare the speed of convergence of the algorithm, as provided by different entanglers, using three blocks in the circuit. The results are collected in Table I. We notice that different placement of CX gates (shown in Table VII) in the entangler block leads to different convergence properties (see Table III). Hence, we scrutinize five selected entanglers (Entanglers 1–5 in Table I) that exhibit similar overall speed of convergence and are constructed with two to three two-qubit gates.

Based on our observations for the basic plaquette, we investigate Hamiltonians describing larger systems composed of adjacent triangular plaquettes, as shown in Figs. 1(b)–1(f). We evaluate whether the performance of selected entangling blocks is preserved, i.e., whether we get close to the ground state with similar statistical accuracy. Our investigation is limited to selected entanglers, that are constructed as natural extensions (see Table VII) of entanglers 1–5, such that they span the full qubit register. We display the results of the speed of convergence in Table III for the Hamiltonians $H_{\Delta_2}$, $H_{\Delta_3}$, $H_{\Delta_4^L}$, $H_{\Delta_4^S}$, and $H_{\Delta_4^P}$. One observes that entanglers of type 1 and 2 perform well in all investigated cases, while types 3–5 cannot be scaled up to perform similarly.

### B. Level of entanglement

For a fixed random initial set of angles, we simulate the VQE algorithm 100 times and extract the mean and the variance of the integrated entanglement (see Table IV and Figs. 4 and 5).

The ground-state wave function $|\Psi_g^{H_{\Delta_1}}\rangle$ has zero three-tangles and nonzero concurrence. Hence we expect to detect bipartite entanglement at instances of convergence (see Fig. 4). Notwithstanding, we can create the three-tangle in the early stages of the optimization procedure (before convergence), which needs to gradually disappear when approaching the ground state. We see this behavior in Figs. 4 and 5. Therefore, we examine whether entanglement can be used as a resource for a speedup, even if the ground state is a separable state.

*Separable Hamiltonian*

For the case of the basic plaquette Hamiltonian, the ground state is an entangled state. In this section we extend the analysis to the separable Hamiltonian (5) with the ground state (6) being a product state. This allows us to compare the speed of convergence of different types of Hamiltonians and the role of entanglement in the process of convergence. For the separable Hamiltonian, one can converge to the ground state by applying local operations (single-qubit gates) without entanglers. It is also possible to approach the ground-state energy using blocks composed of two-qubit gates. However, in all cases considered, the presence of entanglement, for this particular case, slows down the convergence (see Fig. 6). For the separable Hamiltonian entanglement is more an obstacle to overcome than a resource allowing faster convergence.

### C. Scaling and accuracy

In this section we report the results for the scaling of the VQE energy errors as a function of the number $D$ of entangler blocks, and of the accuracy with which the VQE parameters, $\vec{\theta}$, can be set in a digital quantum computer.

TABLE IV. Mean and standard deviation of integrated entanglement (concurrence $\bar{C}_{ij}$ and three-tangle $\bar{\tau}_3$) according to Eq. (17) over 100 runs of the VQE algorithm.

| Hamiltonian | Entangler | $\bar{C}_{01}$ | $\bar{C}_{02}$ | $\bar{C}_{12}$ | $\bar{\tau}_3$ |
|---|---|---|---|---|---|
| | Ent. 1 | 0.612±0.063 | 0.606±0.059 | 0.627±0.040 | 0.086±0.082 |
| | Ent. 2 | 0.634±0.027 | 0.617±0.035 | 0.632±0.025 | 0.065±0.044 |
| $H_{\Delta_1}$ | Ent. 3 | 0.553±0.088 | 0.493±0.117 | 0.479±0.167 | 0.267±0.187 |
| | Ent. 4 | 0.589±0.077 | 0.560±0.087 | 0.575±0.060 | 0.166±0.114 |
| | Ent. 5 | 0.580±0.070 | 0.539±0.089 | 0.570±0.104 | 0.182±0.133 |
| | Ent. 1 | 0.018±0.014 | 0.021±0.018 | 0.031±0.016 | 0.007±0.003 |
| | Ent. 2 | 0.036±0.021 | 0.033±0.017 | 0.043±0.025 | 0.007±0.003 |
| $H_{\text{sep}}$ | Ent. 3 | 0.016±0.014 | 0.015±0.008 | 0.008±0.010 | 0.006±0.003 |
| | Ent. 4 | 0.015±0.008 | 0.014±0.011 | 0.028±0.018 | 0.004±0.002 |
| | Ent. 5 | 0.015±0.007 | 0.014±0.011 | 0.033±0.057 | 0.004±0.003 |

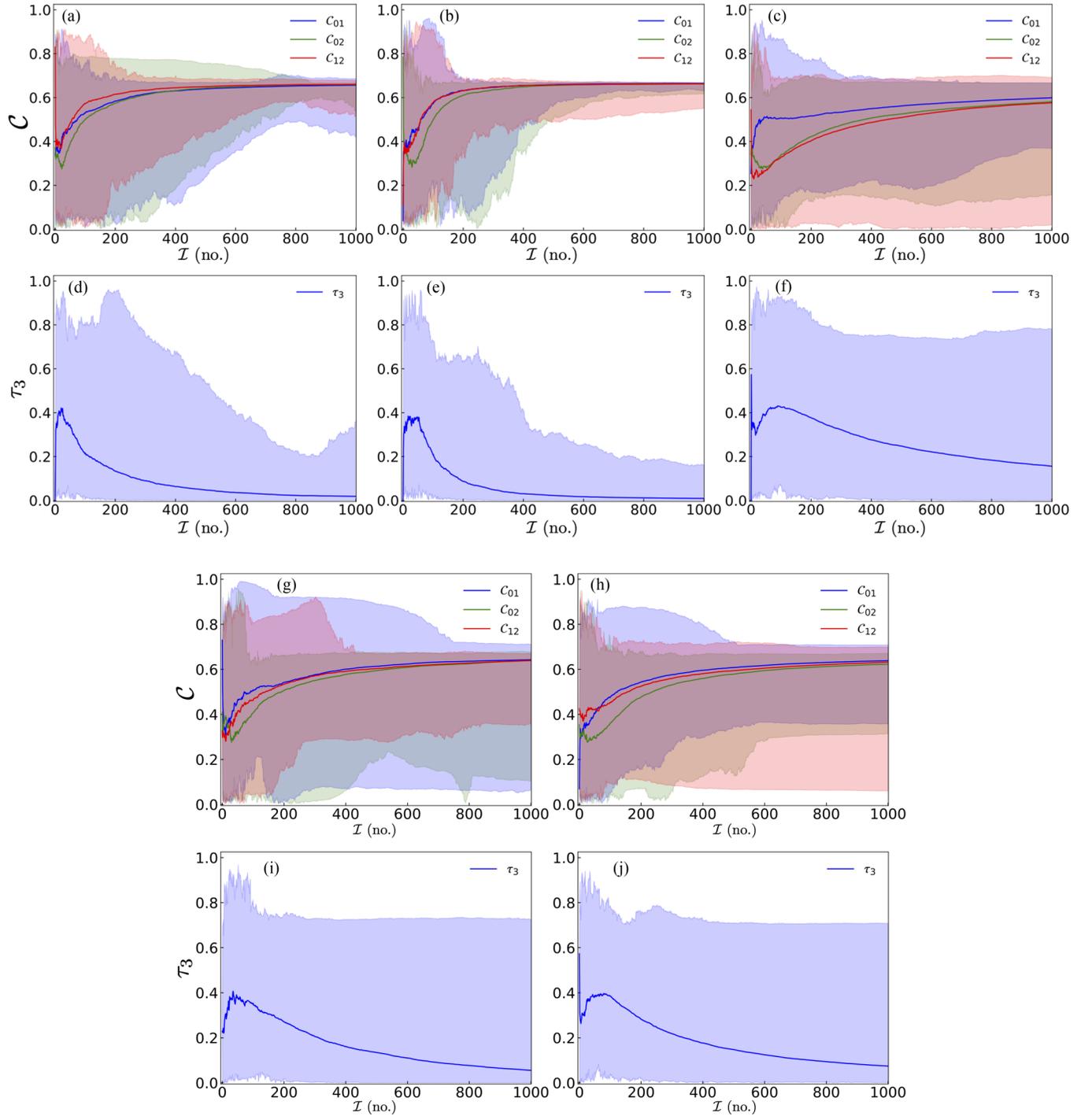

FIG. 4. Concurrence (a)–(c), (g), (h) and three-tangle (d)–(f), (i), (j) present in the trial wave functions during the optimization process for the fermionic triangle Hamiltonian $H_{\Delta_1}$. Solid lines represent average values for both three-tangle and concurrence, while shades show the region between maximal and minimal values of entanglement measures obtained in 100 different runs of the VQE. Subplots correspond to different entanglers (a), (d) Ent. 1, (b), (e) Ent. 2, (c), (f) Ent. 3, (g), (i) Ent. 4, and (h), (j) Ent. 5. Each computation uses three blocks, 100 calibration steps (not displayed), and $\mathcal{I} = 1000$ iterations of the SPSA optimization scheme.

For the description of the most general state in an $N$-qubit system one needs $2^N$ parameters, which is the size of the corresponding Hilbert space. On the other hand, the total number of variational parameters scales linearly with $D$. A large number of variational parameters hence induces a large circuit depth, posing severe challenges for the implementation of the VQE algorithm in NISQ devices. According to the SK theorem (see Sec. III A 2), we can, however, achieve a good approximation of the ground-state solution within an energy error $\epsilon$ using a sequence of length $O[\log_{10}^c(1/\epsilon)]$ of $SU(2^N)$ operations. To estimate the scaling exponent $c$, we performed a series of VQE calculations for the Hamiltonian of the lattice in Fig. 1(d) using $D = 1, \ldots, 12$ entangler blocks of type 1 (see Table I). The convergence of the VQE energy error $\varepsilon$, as a function of $D$ is given in Fig. 7. The fit to the function $\log_{10}^c(1/\epsilon)$ gives a value of $c = 1.31 \pm 0.13$, which is indeed smaller than the limit value of 4 predicted by the SK theorem. The smaller the value of $c$, the shorter the sequence of $SU(2^N)$ operations to achieve an energy accuracy of $\epsilon$. Note that after $D = 5$ the energy error becomes smaller than $\leqslant 5.0 \times 10^{-2}$ (shaded blue area), which corresponds to the limiting value that can be achieved using a maximum of $3 \times 10^4$ SPSA steps for the classical optimizer. In fact, for $D > 5$ we observe a constant value of $\varepsilon$ for the entire range considered (blue line in Fig. 7).

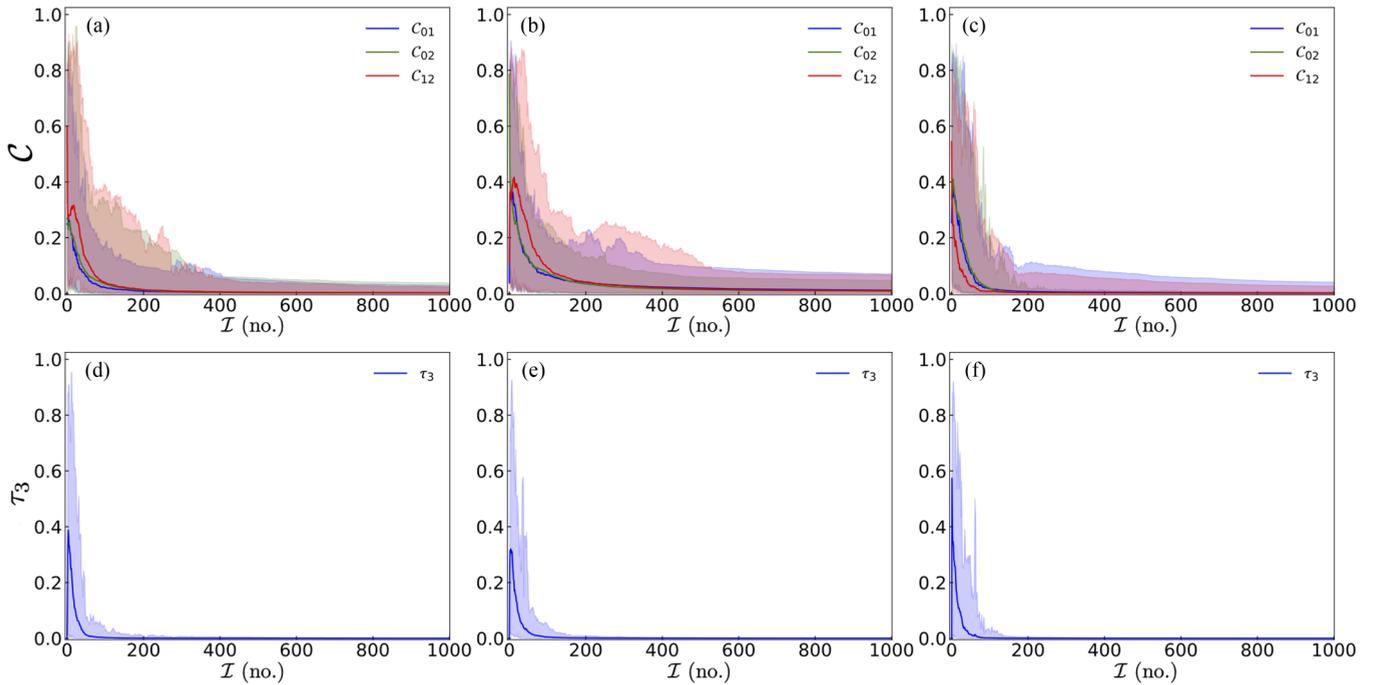

FIG. 5. Concurrence (top panel) and three-tangle (bottom panel) present in the trial wave functions during the optimization process for the separable Hamiltonian. Solid lines represent average values for both three-tangle and concurrence, while shades represent the region between maximal and minimal values of entanglement measures obtained in 100 different runs of the VQE. Subplots correspond to different entanglers chosen from Table I: (a), (d) Ent. 1, (b), (e) Ent. 2, and (c), (f) Ent. 3. Each computation uses three blocks, 100 calibration steps (not shown), and $\mathcal{I} = 1000$ iterations of the SPSA optimization scheme.

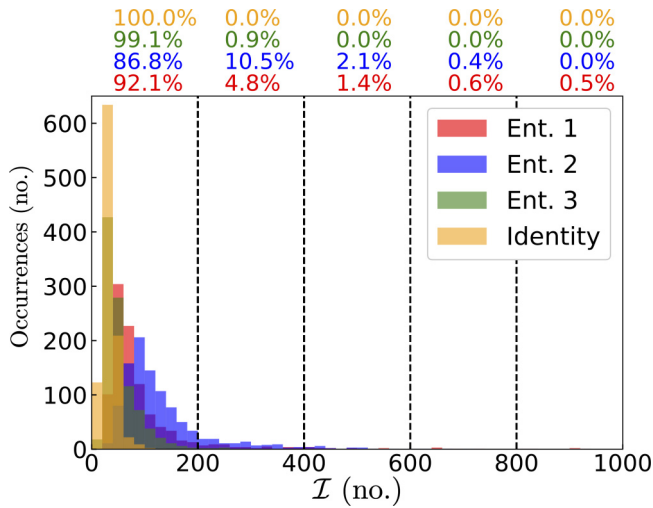

FIG. 6. Convergence statistics of 1000 runs of the VQE with three blocks for the separable Hamiltonian $H_{\text{sep}}$. We call the algorithm converged when the algorithm converges with an error of not more than 2% of the ground-state energy. Each simulation is done with 100 calibration steps and $\mathcal{I} = 1000$ iterations. The plot depicts the entanglers 1 (red), 2 (blue), 3 (green), and 0 (identity) used as blocks. On top of the plots, we show the percentage of runs which converge within intervals of 200 iterations.

In addition to the dependence on the number of blocks, it is also worth investigating the dependence of $\varepsilon_e$ defined as

$$\varepsilon_e = \left| E_{\text{opt}} - E_{\text{VQE}}^{\text{appr}} \right| \quad (18)$$

on the number $D$ of VQE blocks. Also the number of digits of the parameter precision (e.g., the precision of the gate angles) influences $\varepsilon_e$. In Eq. (18), $E_{\text{opt}}$ is the lowest energy, optimized by the VQE, which is obtained using double precision for the qubit parameters (i.e., 72 classical bits). In fact, current hardware for NISQ computing can only achieve a finite digit precision for the setting of the qubit rotations. This introduces a coarse graining of the accessible Hilbert space, allowing approximate solutions only. For every choice of the precision, we first collapsed the "exact" qubit angles by rounding to the corresponding closest approximate value. In this way, the state

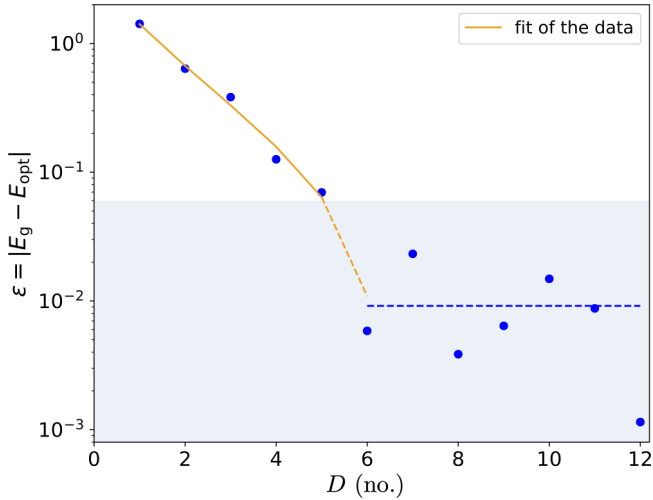

FIG. 7. Dependence of the accuracy of the energy, optimized by the VQE, on the number of entangler blocks $D$, on a log-lin scale. The results correspond to the ground-state energy of the Fermi-Hubbard model described by the lattice (d) of Fig. 1, and are obtained using the entangler block 1 of Table I. For a number of blocks between 1 and 5, the results follow the behavior described by the SK theorem (see Sec. III A 2) with a coefficient $c = 1.31 \pm 0.13$ (orange curve). The threshold value of $5.0 \times 10^{-2}$ (shaded area) defines the maximum accuracy that can be achieved with the VQE algorithm using a maximum of $3 \times 10^4$ SPSA iterations. For $D > 5$ the points show a constant trend (blue dashed line).

vector generated by the same VQE circuit "collapses" to

$$|\psi(\vec{\theta})\rangle \rightarrow |\psi(\vec{\theta}_{\text{DP}})\rangle \qquad (19)$$

where $\vec{\theta}_{\text{DP}}$ is the approximated set of angles with decimal precision. The corresponding approximate energy is then evaluated as

$$E_{\text{VQE}}^{\text{appr}} = \langle \psi(\vec{\theta}_{\text{DP}})|H|\psi(\vec{\theta}_{\text{DP}})\rangle. \qquad (20)$$

Even though the differences among all entanglers is not large, we observe a faster error reduction for entanglers 1 and 2, which also provide faster convergence (see Table V). Interestingly, we observe that in order to achieve an eight digits precision for the final ground-state energy only a modest accuracy in the angle setting is required (DP ≈ 4). This result is particularly relevant for calculations performed on quantum hardware, where current technological restrictions are limiting the accuracy with which gate angles can be set.

TABLE V. Speed of convergence for five different entanglers.

| | Convergence within 2% for the number of iterations (%) | | | | | |
|---|---|---|---|---|---|---|
| Ent. | 1–200 | 201–400 | 401–600 | 601–800 | 801–1000 | Total |
| 1 | 71.4 | 18.5 | 4.6 | 2.0 | 0.2 | 96.7 |
| 2 | 72.6 | 20.8 | 4.3 | 0.9 | 0.2 | 98.8 |
| 3 | 25.3 | 32.7 | 15.0 | 8.4 | 1.0 | 82.4 |
| 4 | 29.7 | 34.9 | 14.5 | 4.6 | 0.4 | 84.1 |
| 5 | 23.3 | 39.6 | 16.3 | 6.0 | 0.4 | 85.6 |

## V. CONCLUSIONS

In this paper, we investigated the properties of the variational quantum eigensolver algorithm for the determination of the ground-state energy of noninteracting Fermi-Hubbard models, through a systematic analysis of a series of trial wave functions and quantum circuits. In particular, we focused on the analysis of a three-site Hamiltonian $H_{\Delta_1}$, for which we additionally created a separable Hamiltonian $H_{\text{sep}}$ with the same spectrum but different eigenstates. To assess the exact physical properties, all our numerical calculations were performed using high-precision simulations of the quantum circuits on classical hardware, i.e., without including any type of noise sources that occur in NISQ calculations.

Particular care was given to the study of the amount of entanglement created during the optimization process and its impact on the convergence of the algorithm. We found that a variety of circuits ensure the convergence of the algorithm towards the correct ground state, generating the needed amount of entanglement. However, while the nature of the circuit clearly determines the level of entanglement that can be achieved, the path followed by the evolving state vector in Hilbert space, together with the corresponding entanglement profile, also depend on the employed optimization routine (here always SPSA, see Sec. III.C). We observed that those entanglers which allow the optimization routine to create appreciable bipartite entanglement alone perform, on average, better than the ones creating both bipartite and tripartite entanglement in the course of the target state search. Since the entanglement of the ground state of the basic plaquette

TABLE VI. Gate action expressed either by a unitary matrix or by its action on a state vector.

| Circuit | Name | Matrix |
|---|---|---|
| | CX | $\begin{pmatrix} 1 & 0 & 0 & 0 \\ 0 & 1 & 0 & 0 \\ 0 & 0 & 0 & 1 \\ 0 & 0 & 1 & 0 \end{pmatrix}$ |
| | CZ | $\begin{pmatrix} 1 & 0 & 0 & 0 \\ 0 & 1 & 0 & 0 \\ 0 & 0 & 1 & 0 \\ 0 & 0 & 0 & -1 \end{pmatrix}$ |
| | SWAP | $\begin{pmatrix} 1 & 0 & 0 & 0 \\ 0 & 0 & 1 & 0 \\ 0 & 1 & 0 & 0 \\ 0 & 0 & 0 & 1 \end{pmatrix}$ |
| | iSWAP | $\begin{pmatrix} 1 & 0 & 0 & 0 \\ 0 & 0 & i & 0 \\ 0 & i & 0 & 0 \\ 0 & 0 & 0 & 1 \end{pmatrix}$ |
| | Toffoli | $\begin{pmatrix} \mathbb{1}_6 & 0 & 0 \\ 0 & 0 & 1 \\ 0 & 1 & 0 \end{pmatrix}$ |
| QFT | Quantum Fourier Transform | $U_{QFT}|x\rangle = \frac{1}{\sqrt{2^3}} \sum_{k=0}^{7} e^{\frac{2\pi i}{8}kx} |k\rangle$ |

Hamiltonian is of bipartite kind, it appears suggestive that creation of the wrong type of entanglement is detrimental for the algorithm's convergence. Consistently, in the case of the Hamiltonian with a separable ground state, any type of entanglement is decreasing the efficiency of the VQE algorithm, slowing down the convergence process.

Therefore, one needs to proceed cautiously when referring to entanglement as a resource for potential quantum speedup, and always take into account the physical nature of the problem under study. Entanglement between arbitrary or not suitable parties may hamper the convergence process.

Within the model Hamiltonians considered in this paper, we found that entanglers built from CX gates provide faster convergence than the ones based on CZ or ISWAP gates. This is a promising result, since CX gates are native to implement on many available NISQ quantum devices (e.g., IBM QX). Additionally, entanglers composed of fewer gates potentially perform better on real devices because of the limited impact of the gate errors and fidelities on the final results. For these reasons, we argue that the "type 1" entanglers are the entanglers of choice for the noninteracting Fermi-type models (of all investigated dimensionalities) described in this paper. In addition, one has to bear in mind that the efficiency of the VQE depends on the number of blocks the circuit is built from.

The convergence of the VQE energies as a function of the number of entangler blocks (i.e., of the number of parametrized gate operations) was assessed for a six-qubit Hamiltonian corresponding to the lattice in Fig. 1. For a particular system choice, we showed that the accuracy of the ground-state energies follows the behavior predicted by the SK theorem, with an exponent $c \approx 1.3$. While not generally applicable to all other Hamiltonians and lattice geometries, this result confirms that the VQE algorithm can reproduce energies with accuracy $\varepsilon$ using a number of gate operations that scales like $O[\log_{10}^c(1/\varepsilon)]$. Further analysis is needed to demonstrate the validity of these results for the more general Fermi-Hubbard models with intrastate Coulomb electronic repulsion.

TABLE VII. Scaling of entanglers.

| | $\Delta_2$ | $\Delta_3$ | $\Delta_4^{P,L,S}$ |
|---|---|---|---|
| Ent.1 | 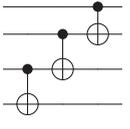 | 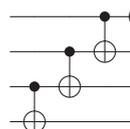 | 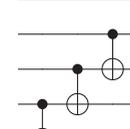 |
| Ent.2 | 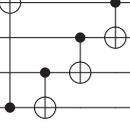 | 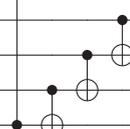 | 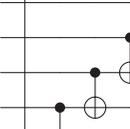 |
| Ent.3 | 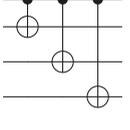 | 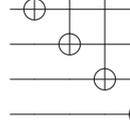 | 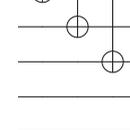 |
| Ent.4 | 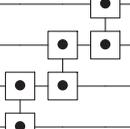 | 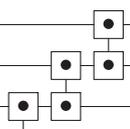 | 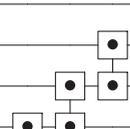 |
| Ent.5 | 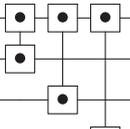 | 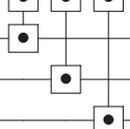 | 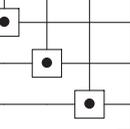 |

## ACKNOWLEDGMENTS

The authors would like to thank Sergey Bravyi, Antonio Mezzacapo, Kristan Temme, Anton Robert, Stefan Woerner, Pauline Ollitrault, Igor Sokolov, Nikolaj Moll, and Abhinav Kandala for useful discussions. F.W. was partially supported by the Polish Ministry of Science and Higher Education program "Mobility Plus" through Grant No. 1278/MOB/IV/2015/0 and thanks NASA Ames Research Center for support. P.B. and I.T. acknowledge support from the Swiss National Science Foundation through Grant No. 200021-179312. A.W. is indebted to the German National Academic Foundation and to the Konrad Adenauer Foundation.

## APPENDIX A: MATRIX REPRESENTATION OF QUANTUM GATES

In Table VI, we state the representation of the quantum gates as unitary matrices.

## APPENDIX B: SCALING OF ENTANGLER BLOCKS

The entangler blocks used for the triangles consisting of more than three sites are depicted in Table VII.

ENTANGLEMENT PRODUCTION AND CONVERGENCE … PHYSICAL REVIEW A **102**, 042402 (2020)